\documentstyle[aps,prb,epsfig]{revtex}
\begin{document}

\twocolumn[
 \hsize\textwidth\columnwidth\hsize
 \csname@twocolumnfalse\endcsname

\draft
\title{The excitation spectrum of mesoscopic proximity structures}
\author{S. Pilgram$^1$, W. Belzig$^2$, and C. Bruder$^1$}

\address{$^1$Departement Physik und Astronomie, Universit\"at Basel,
Klingelbergstr.~82, CH-4056 Basel, Switzerland\\
$^2$Theoretical Physics Group, Delft University of Technology,
Lorentzweg~1,NL-2628 CJ Delft, The~Netherlands}

\maketitle

\begin{abstract}
We investigate one aspect of the proximity effect, viz., the local density
of states of a superconductor-normal metal sandwich. In contrast to earlier
work, we allow for the presence of an arbitrary concentration of impurities
in the structure. The superconductor induces a gap in the normal metal
spectrum that is proportional to the inverse of the elastic mean free path
$l_{\text{N}}$ for rather clean systems. For a mean free path much shorter
than the thickness of the normal metal, we find a gap size proportional to
$l_{\text{N}}$ that approaches the behavior predicted by the Usadel equation
(diffusive limit). We also discuss the influence of interface and surface
roughness, the consequences of a non-ideal transmittivity of the interface,
and the dependence of our results on the choice of the model of impurity
scattering.
\end{abstract}

]

\section{Introduction}

A normal metal in good metallic contact to a superconductor acquires
superconducting properties like infinite conductance and the Meissner
effect (see \cite{superlattices} and references therein). This
so-called {\em proximity effect} has been extensively studied in the
past decade mainly in the two limiting cases of fully ballistic
propagation and diffusive motion. Whereas the first may be realized
experimentally in two-dimensional electron gases, the latter is
realized in the most common samples made of semiconducting or
structured metallic films.

Although the ballistic and the diffusive cases provide useful bounds,
real-world mesoscopic samples are often in the intermediate regime.
For instance, the diamagnetic response of mesoscopic proximity
cylinders (superconducting wires covered by a normal metal) has
attracted a lot of experimental \cite{mota,motanew,mueller} and
theoretical \cite{degennes,zaikin,nagano,belzigbruderfauchere}
interest. It turned out that the experimental results could only be
understood by considering intermediate impurity concentrations
\cite{mueller}. Previous theoretical analyses in the clean limit
\cite{zaikin} and the dirty limit \cite{degennes,nagano} were not able
to explain the experimental results quantitatively. Only recently it
was shown that a qualitative different behavior emerges in the regime
of intermediate impurity concentration \cite{belzigbruderfauchere}.
With the help of this calculation a quantitative agreement with the
experimental results could be obtained \cite{mueller} and it was
possible to determine the mean free path $l_{\text{N}}$.

The precise determination of the degree of disorder is relevant for a
characterization of the samples. It may serve as a basis for the
understanding of the the experimental observation of a low-temperature
paramagnetic reentrance effect \cite{mota,motanew}. This has already
stimulated theoretical suggestions that orbital currents might lead to
a paramagnetic contribution \cite{reentrance1,reentrance2}. These
currents depend on the degree of disorder in the normal metal and
further investigations are necessary to quantify the influence of
disorder.

In the present work, we want to focus on another aspect of the
proximity effect, viz., on the change of the excitation spectrum of a
normal metal connected to a superconductor. In particular, we are
interested in the changes of the spectrum due to disorder.

In this paper, we report on a comprehensive study on the density of
states in a moderately dirty proximity sample. The paper is organized
as follows.  In the next section we introduce the geometry and the
parameters of our model.  In Sec.~\ref{sec:bulkdisorder} we present
results for the density of states for different disorder
concentrations and models. Finally, we discuss the effect of rough
surfaces and rough interfaces in Sec.~\ref{sec:surfacedisorder}.

\section{Geometry and model}
\label{sec:model}

\begin{figure}
\begin{center}
\leavevmode
\psfig{file=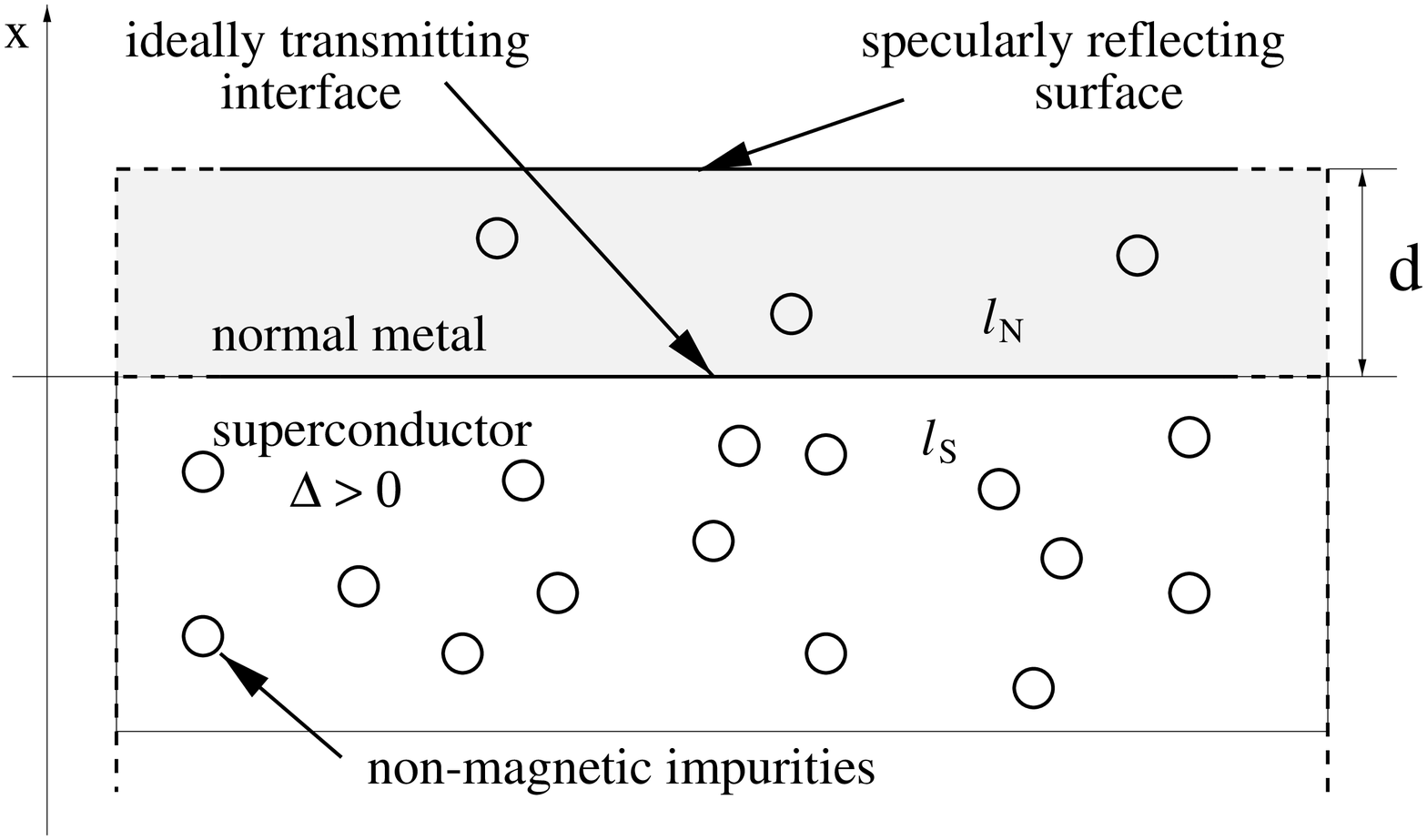,width=0.9\columnwidth}
\caption{
  Geometry of the SN interface. $l_{\text{N}}$ ($l_{\text{S}}$) is the
  elastic mean free path of the normal (superconducting) part. In the
  first part of the paper, the interface is supposed to be ideal and
  the surface specularly reflecting.}
\label{geometry}
\end{center}
\end{figure}

The sample geometry that we have in mind is shown in
Fig.~\ref{geometry}: we consider a slab geometry in which a normal
metal layer of thickness $d$ is connected by an ideal interface to a
superconductor. The outer surface of the sandwich is supposed to be
specularly reflecting. In the following, we will discuss the local
density of states (LDOS) $N(E,x)$ of this structure for a variety of
physical situations. The quasiclassical formulation of
superconductivity is most suitable to calculate the LDOS. To this end
we have solved the real-time version of the Eilenberger equation
\cite{eilenberger,larkin}
\begin{equation}
 \label{eilenberger_equations}
 -{\bf v_F \nabla}\hat{g}({\bf v_F},{\bf r},E) =
 \left[ -iE\hat{\tau}_3 + \Delta({\bf r})\hat{\tau}_1 
 + \hat{\sigma}({\bf r},E),\hat{g}({\bf v_F},{\bf r},E) \right] 
\end{equation}
for the quasiclassical $2\times 2$ matrix Green's function $\hat g$
numerically (see \cite{superlattices} and Appendix \ref{numerics} for
additional details of this method). Here and in the following, $\hbar$
is set to one except for the final results. In
Eq.~(\ref{eilenberger_equations}), $\Delta$ is the pair potential and
$\hat{\sigma}$ the impurity self-energy which has to be determined
self-consistently by a scattering matrix equation.  The Pauli matrices
$\hat{\tau}_i$ are used as a basis for the $2\times 2$ matrix
equation, ${\bf v_F}$ is an arbitrarily oriented unit vector times the
Fermi velocity. The quasiclassical Green's function is normalized
according to
\begin{equation}
\label{normalization}
\hat{g}^2({\bf v_F},{\bf r},E) = 1\; .
\end{equation}
In principle, the pair potential $\Delta$ has to be determined
self-consistently in the superconductor, whereas it vanishes by
definition in the normal metal. Since we are mainly interested in the
properties of the normal metal in the limit $d\gg \hbar
v_{\text{F}}/\Delta$, the spatial dependence of the pair potential in
the superconductor plays no role. Thus, we approximate the pair
potential by a step function $\Delta({\bf r})=\Delta\theta(-x)$.

The effect of impurities give rise to the self-energy $\hat\sigma
({\bf r})$.  In general the impurity self energy can be found from a
solution of a t-matrix equation \cite{bucholtz} In our particular case
this equation can be solved and presented as
\begin{equation}
 \label{unitarity-self-energy}
 \hat{\sigma}({\bf r},E) = 
 \frac{n_{\text{imp}}({\bf r})v^2({\bf r})
 N_0 \langle\hat{g}({\bf r},E)\rangle}{
 1 + v^2({\bf r})N_0^2\langle\hat{g}({\bf r},E)\rangle^2}\; .
\end{equation}
This impurity self-energy contains two parameters, the impurity
concentration $n_{\text{imp}}$ and the strength of the scattering
potential $v({\bf r})$ (that may be spatially dependent if there are
different impurities in different parts of the sample). Additionally,
the normal metal-density of states at the Fermi level $N_0$ enters
into Eq.~(\ref{unitarity-self-energy}).

In the rest of the paper we will mostly use the Born approximation
valid in the limit of weak scattering, where the denominator in
Eq.~(\ref{unitarity-self-energy}) can be neglected:
\begin{equation}
 \label{eq:born-self-energy}
 \hat{\sigma}({\bf r},E) = \frac{1}{2\tau_{\text{imp}}({\bf r})}
 \langle\hat{g}({\bf r},E)\rangle\; .
\end{equation}
Here, we have introduced the elastic scattering time
$\tau_{\text{imp}} = (2n_{\text{imp}}({\bf r})v^2({\bf r})N_0)^{-1}$
which is related to the mean free path $l = v_{\text{F}}
\tau_{\text{imp}}$. Since we will also consider inhomogeneous impurity
distributions (on a mean level), we still allow for a spatial
dependence of elastic scattering time and mean free path.

The set of equations
(\ref{eilenberger_equations})-(\ref{eq:born-self-energy}) has been
solved numerically. The integration of Eq.~\ref{eilenberger_equations}
was performed using the Riccati parametrization \cite{schopohl} (see
Appendix~\ref{parametrization}). The self-energies had to be
determined self-consistently for all energies in an iterative scheme.
Finally, the local density of states is given by
\begin{equation}
 \label{eq:local-density-of-states}
 N(E,{\bf r})=N_0 \text{Re}\langle\frac12 {\text{Tr}}\hat\tau_3
 \hat{g}({\bf r},E)\rangle \; .
\end{equation}

\begin{figure}
\begin{center} 
\leavevmode
\psfig{file=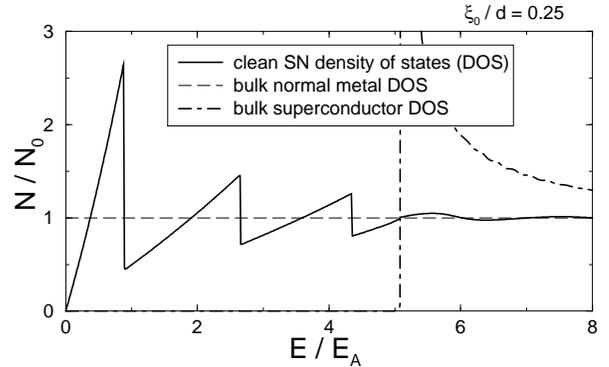,width=0.9\columnwidth}
\caption{Density of states of a clean normal metal connected to a
  superconductor. In the clean case, $N(E,x)=N(E)$, i.e., there is no
  dependence on the spatial coordinate. The coupling to the
  superconductor leads to a linear suppression of the DOS at the Fermi
  energy. Here, we chose $\Delta=5 E_{\text{A}}$.}
\label{cleandos}
\end{center}
\end{figure}

Before we discuss our results, let us briefly summarize the known
spectral properties in the limiting cases of clean and dirty limit.
The case of a clean normal metal was already discussed in the 60's:
the LDOS is independent of the location in the normal metal, vanishes
at the Fermi energy and rises linearly close to it
\cite{DOSLIN,mcmillan}. Its peculiar zig-zag from is shown in
Fig.~\ref{cleandos}. The characteristic energy determining the jumps
in the spectrum is the Andreev energy $E_{\text{A}}$. This
characteristic energy follows from a semi-classical quantization
condition for the Andreev bound states. According to this rule we have
to add the phases accumulated along a trajectory burying a bound
electron-hole state. This phase difference is given by twice the phase
shift of an Andreev reflection at the superconductor, which at
$E\ll\Delta$ is given by $\pi/2$. On the path through the normal metal
both electron and hole accumulate an additional shift of $2\times 2 E
d/v_{\text{F}}$, proportional to twice the time spent in the normal metal.
Adding all contributions to the semiclassical phase and requiring it
to be an integer multiple of $2\pi$ leads to the characteristic energy
of the lowest level
\begin{equation}
 \label{eq:andreev-energy}
 E_{\text{A}}=\frac{\hbar\pi v_{\text{F}}}{4d}\; .
\end{equation}
The subsequent levels are (approximately) equidistant with level
spacing $2E_{\text{A}}$ as can be seen from Fig.~\ref{cleandos} (the
deviations from the ideal level spacing are caused by the finite value
of $\Delta \approx 5E_{\text{A}}$).

Another well-known result on the spectrum has been obtained in the
dirty (diffusive) limit \cite{golubov,belzigdos}. In this case, the
LDOS is characterized by a minigap in the spectrum that is of the
order of the Thouless energy $E_{\text{Th}}=\hbar D/d^2$, here $D$ is
the diffusion constant of the normal metal and $d$ its thickness. The
LDOS of a mesoscopic superconductor-normal metal sample was determined
experimentally \cite{pothier:96} in the presence of a magnetic field,
and our theory \cite{belzigdos} led to a satisfactory understanding of
those experimental results.

The induced minigap has also been discussed by field-theoretic means,
see \cite{altlandreview}. The relation of this gap to quantum chaos is
discussed in Refs. \cite{chaos}.

A qualitative picture of the formation of the minigap follows from
considering the Andreev bound states in the normal layer. An electron
and hole traversing a diffusive trajectory in the time reversed
direction are transformed into each other by Andreev reflection when
hitting the superconductor. The coherent superposition of two
subsequent reflections result in a bound state. The phase shift due to
Andreev reflection is similar as in the clean limit discussed above.
During the motion the two quasiparticle gain an additional phase shift
$2E\delta t$, where $\delta t$ is the time spent in the normal metal
region and $2E$ is the energy difference of electron and hole. In a
diffusive system the maximal time spent in the normal metal is $\sim
2d^2/D$, i.e., twice the time to diffusion time. This upper time limit
results in a lower bound to the energy $\sim \pi D/4d^2$ above which
constructive interference is possible. The minigap is thus expected to
be approximately $ E_{\text{g}}\approx \pi \hbar D/ 4 d^2 = 0.785
\hbar D/d^2$. The numerically exact expression for the minigap in the
dirty case is given by $E_{\text{g}}=0.780\hbar D/d^2$, where the
numerical factor is the solution of a transcendental equation (see
Appendix \ref{dirty minigap}).  This result is in very good agreement
with the estimate given before corroborating the simple picture of
Andreev bound states.

\section{Arbitrary impurity concentration}
\label{sec:bulkdisorder}
How is the linear rise of the LDOS for the clean system transformed
into the minigap in the diffusive system as a function of impurity
concentration? To answer this question, we have solved the real-time
Eilenberger equation including an impurity self-energy of the form
$\hat{\sigma}=\langle \hat g \rangle/2\tau$ (Born approximation). The
impurity self-energy was determined in a self-consistent way.

\begin{figure}
\begin{center}
\leavevmode
\psfig{file=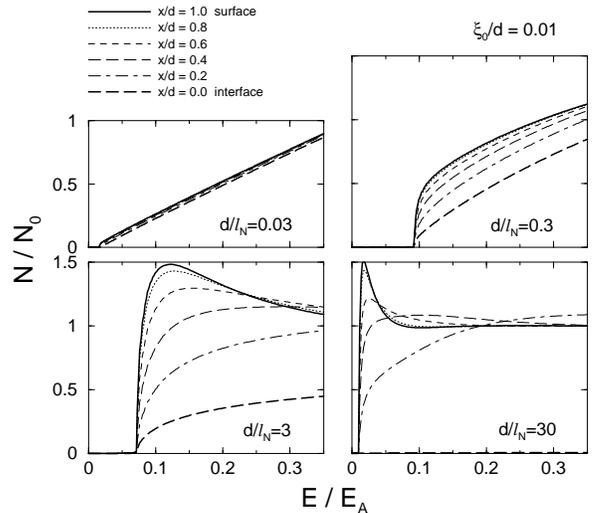,width=0.9\columnwidth}
\caption{
  LDOS $N(E,x)$ for different ratios of mean free path $l_{\text{N}}$
  to normal-layer thickness $d$. The minigap is constant throughout
  the normal metal, but the energy dependence of the LDOS changes with
  location.}
\label{ldos}
\end{center}
\end{figure}

\begin{figure}
\begin{center}
\leavevmode
\psfig{file=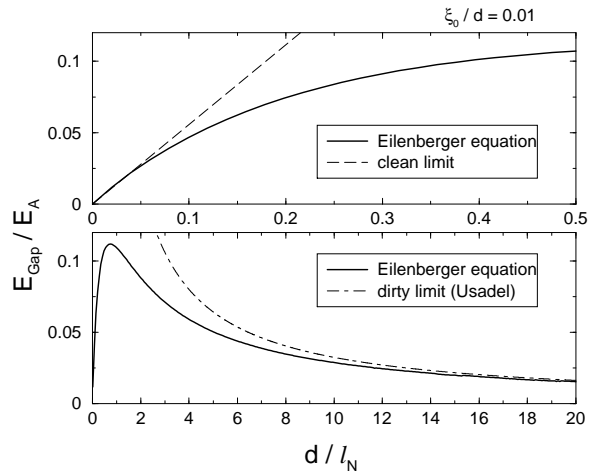,width=0.9\columnwidth}
\smallskip
\caption{Size of the minigap obtained by numerically solving the 
Eilenberger equation.}
\label{gapsize}
\end{center}
\end{figure}

We find \cite{pilgramdiplom,pilgram} that a gap forms at arbitrarily
small impurity concentrations. This is shown in Fig.~\ref{ldos}: even
for values of the elastic mean free path $l_{\text{N}}$ that are $30$
times larger than the normal-layer thickness, the formation of the
low-energy gap is clearly visible. The gap increases with
$1/l_{\text{N}}$, saturates for $l_{\text{N}}\sim d$ and then
decreases again as expected from the dirty-limit theory since $D\sim
v_{\text{F}} l_{\text{N}}$, see Fig.~\ref{gapsize}. The gap does not
depend on the location in the normal metal as can be seen in
Fig.~\ref{ldos}, i.e., it is a global feature. The shape of the LDOS,
i.e., its dependence on energy, however, varies on traversing the
normal layer.

The existence of a minigap is in line with qualitative considerations
given by McMillan \cite{mcmillan}. He argued that -- quite generally
-- the density of states should show a gap of order of the inverse of
the escape time, i.e., the time an electron spends in the normal layer
before being Andreev-reflected.  If we replace the escape time by the
scattering time in the (almost) clean system and by the diffusive
escape time in the dirty system, we obtain the non-monotonous behavior
shown in Fig.~\ref{gapsize}.

\begin{figure}[tb]
 \begin{center}
 \leavevmode
 \psfig{file=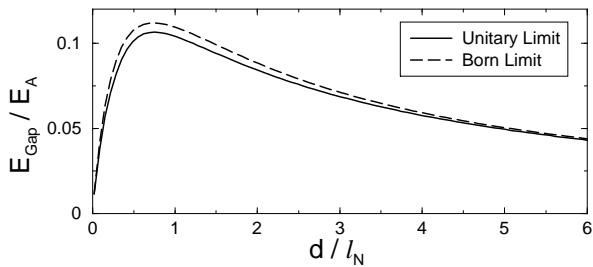,width=0.9\columnwidth}
\smallskip
 \caption{
   Dependence of the induced minigap on bulk disorder for two models
   of impurity scattering, viz., Born and t-matrix approximation. The
   differences are insignificant, i.e., the minigap is stable and is
   not dependent on the choice of the Born approximation.}
 \label{unitary}
 \end{center}
\end{figure}

We have also replaced the Born model for impurity scattering by the
unitary limit which follows from the t-matrix approximation for s-wave
scattering Eq.~(\ref{unitarity-self-energy}) in the limit $ v N_0\gg
1$.  To facilitate a comparison we used for the parameter $n_i/N_0$,
characterizing the strength of the impurity self-energy in the unitary
limit, the value of $1/2\tau$ from the Born approximation.
Figure~\ref{unitary} shows the minigap for both the Born and the
unitary approximation. The minigap is slightly reduced in the
unitary limit, but its functional dependence on the mean free path
is practically unchanged.

\section{Interface and surface roughness}
\label{sec:surfacedisorder}

\begin{figure}
 \begin{center}
 \leavevmode \psfig{file=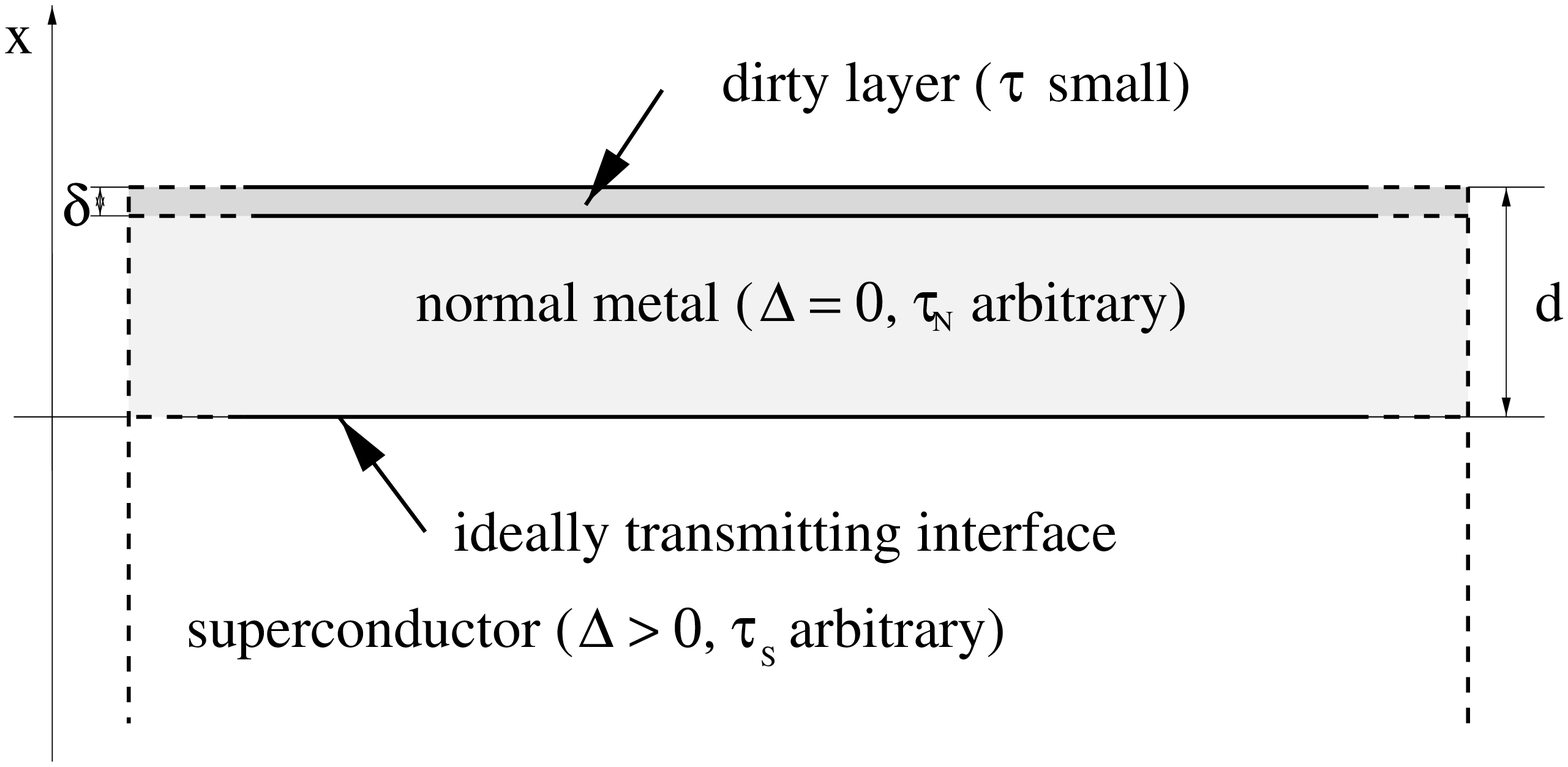,width=0.9\columnwidth}
\smallskip
 \caption{We model a rough surface (or interface) by replacing it by
   a thin dirty layer. The scattering time $\tau$ in this layer and
   its thickness are chosen such as to lead to the proper scattering
   characteristics of a rough surface.}
 \label{modelofdirtylayer}
 \end{center}
\end{figure}

Real-life interfaces and surfaces of typical proximity samples are
rough (for an example, see the photograph in the second paper of Ref.
\cite{mota}). In the quasiclassical language such a roughness will
lead to a mixture of different trajectories, thus smearing out the
singular Andreev bound states on a given path. A convenient way to
include this effect in the quasiclassical formalism is to model the
rough surface by a thin dirty layer which covers the inner side of the
surface, see Fig.~\ref{modelofdirtylayer}. The thickness $\delta$ of
the layer should be taken to be much smaller than all characteristic
lengths of the rest of the system (in in our case this are the
thickness of the normal metal $d$ and its mean free path
$l_{\text{N}}$). The disorder in the layer is included by an impurity
self-energy with a mean free path $l_{\text{layer}}$. At the outer
surface inside the dirty layer specular reflection is assumed and the
Green's functions in the dirty layer are continuously connected to the
ones in the normal metal. Under these conditions the ratio
$l_{\text{layer}} / \delta$ is the only parameter measuring the
roughness of the real-life interface.

Similarly the interface roughness between the normal metal and the
superconductor is modeled by a thin dirty layer residing now at the
interface. The layer is characterized by the same parameters as
before. Now we have continuity of the Green's function on both sides
of the dirty layer.

It is clear that this model for roughness should be taken with care,
since it is by no means microscopically justified. The parameters
characterizing the dirty layer are not related to the real parameters
of the interface. It is however clear that this model covers the
essential properties of a realistic rough surface, i.e. it couples
classical trajectories which would be uncoupled for a specular
interface. Thus it leads to an averaging over spectral quantities from
different regions of the system in the same way as a realistic system
does on average. One should however keep in mind, that we can only
determine averages over many characteristic lengths on the roughness
in this way. E.~g. in small cavities with a characteristic length of
roughness which is comparable to the size of the system, such an
averaging is not appropriate and fluctuations may become important.

\subsection{Surface disorder}
\label{sec:sufrace-disorder}

The result of a calculation for a rough outer surface is shown in the
left plot of Fig.~\ref{roughsurface}. We have kept the roughness of
the surface fixed in this plot and varied the bulk disorder. Only the
LDOS at the outer surface is plotted. In our example, the roughness
parameter is such that there are on average two scattering events in
the dirty layer, i.e., the outer surface is definitely non-specular.
The induced minigap is not very susceptible to the presence of the
surface roughness as follows from a comparison with Fig.~\ref{ldos}.
Using McMillan's argument that the minigap should be inversely
proportional to the escape time, we can qualitatively understand this
behavior. The effect of the scattering by the rough surface leads to a
reduction of the number of the shorter trajectories, but the length of
the shortest trajectory itself is not changed. This can be most
clearly seen for an intermediate bulk disorder $l\approx 3d$, where
the sharp increase at the minigap for the specular surface is nearly
absent. The rough surface also leads to a change in the spatial form
of the density of states, which is not shown in the plot.

We remark in passing that surface roughness {\em without} bulk
disorder will not lead to the formation of a minigap, since there will
be no upper cutoff for the trajectory lengths in this case.
Nevertheless it influences the energy dependence of the LDOS, by
virtue of a similar effect as mentioned previously. Due to the
reduction of the number of the longer trajectories spectral weight is
shifted to higher energies. This can already be seen from the solid
curve in the left plot Fig.~\ref{roughsurface}, where the linear
energy dependence in the case of specular reflection has turned into
some weaker energy dependence.

\subsection{Interface disorder}
\label{sec:interface-disorder}

In many experimental situation the interface between a normal metal
and a superconductor will be nonideal in the sense that either
transport through the interface is not along classical trajectories or
the electrons are partially reflected. This can be due to either an
oxide layer or an alloy in the interface region, and/or due to
differences in effective masses or Fermi velocities between the
superconductor and the normal metal. The generic properties of these
disordered interfaces can be included in the quasiclassical formalism
by a thin dirty layer located at the interface. This model will also
include the effect of a finite reflection at the interface. Since it
is a well know universal property of disordered conductors to have a
bimodal distribution of transmission eigenvalues (see e.~g.
\onlinecite{beenakker:97}), we believe that the model of such an
disordered layer will cover most of the characteristics of rough
contacts.

\begin{figure}
\begin{center}
 \leavevmode
 \psfig{file=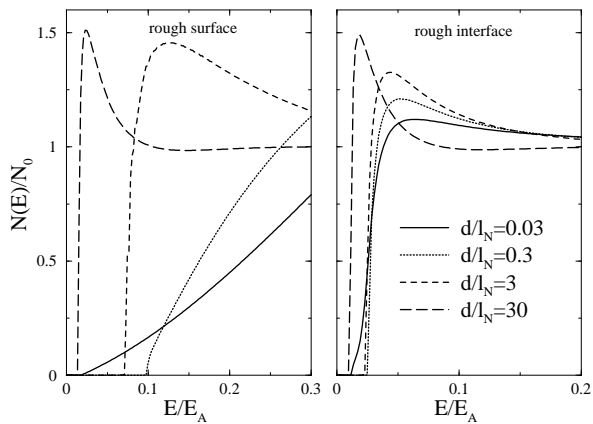,width=0.9\columnwidth}
 \smallskip
 \caption[]{
   LDOS at the outer surface ($x=d$) under the influence of surface
   disorder (left plot) or interface disorder between superconductor
   and normal metal (right plot). The thickness of the layer is chosen
   to be $\delta=10^{-5}d$ and the mean free path in the layer is
   $l_{\text{layer}}=\delta/2$. The different curves in each plot are
   for different degrees of disorder in the normal metal as indicated
   in the right plot. A comparison with Fig.~\protect\ref{ldos} shows
   that the size of the minigap is practically conserved in the left
   plot, whereas the non-ideal interface leads to a decreased minigap
   in the right plot.}
\label{roughsurface}
\end{center}
\end{figure}

\begin{figure}[tb]
\begin{center}
\leavevmode
\psfig{file=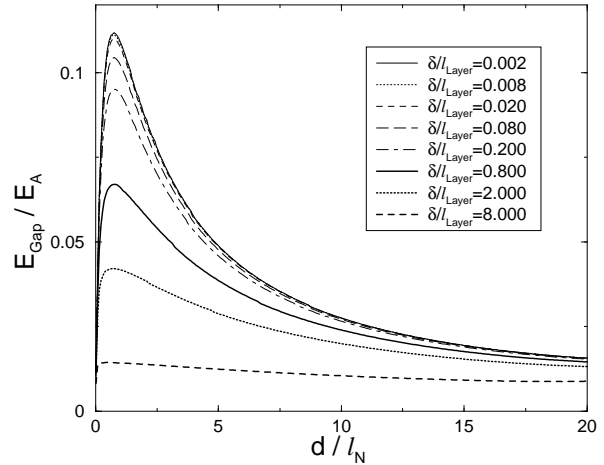,width=0.9\columnwidth}
\smallskip
\caption{Influence of interface roughness (parametrized by
  $\delta/l_{\text{layer}}$) on the induced minigap in the normal
  layer. In contrast to surface roughness, interface roughness leads
  to a pronounced suppression of the minigap. In particular roughness
  parameters larger than $1$ lead to a strong suppression of the
  minigap.  }
\label{roughinterface}
\end{center}
\end{figure}

In the right plot of Fig.~\ref{roughsurface} we show the effect of a
thin dirty layer at the interface between superconductor and normal
metal. The interface is chosen to have a mean free path
$l_{\text{N}}=\delta/2$, similar to the previous case of the rough
surface. This is supposed to mimic a rough interface between the
metals with equal Fermi velocities. We find a significant reduction of
the minigap. This is clearly seen from comparison with the left plot
of Fig.~\ref{roughsurface} and shown as a function of the interface
roughness parameter in Fig.~\ref{roughinterface}. For roughness
parameters $\delta/l_{\text{layer}}$ smaller than $\approx 1$ the
reduction of the minigap is weak. If this parameter is of the order of
or larger than $1$ the minigap is strongly decreased.

This behavior can be understood by looking at the reflection
properties of the rough interface. If the roughness parameter is
smaller than $1$ the reflectivity will be roughly given by $R\approx
\delta/l_{\text{layer}}$, i.e., the probability for a particle to be
reflected is given by the ratio of the mean time spent in the layer to
the scattering time. As a rough estimate of the effect of the finite
reflectivity $R>0$ for the minigap, we use the fact that the mean time
that an electron spends in the normal metal will be increased by a
factor $1+R+R^2+..=1/T$, where $T$ is the transmission eigenvalue of
the interface. Hence, the minigap will be reduced by a factor
$T\approx(1-\delta/l_{\text{layer}})$. The influence of finite
reflection coefficients in the clean system was also discussed in
\cite{nagai}.

If the interface is strongly disordered, i.e., the roughness parameter
is larger than $1$, it behaves more like a disordered metal, having a
bimodal distribution of transmission eigenvalues. Most of the
transmission eigenvalues are close to zero and a few are close to
unity allowing for Andreev reflection. The ratio of the number of
closed to the number of open trajectories is roughly given by
$\delta/l_{\text{layer}}$. Thus, an electron typically is normally
reflected by the dirty layer many times until it has the chance to be
Andreev reflected. This strongly enhances the length of the
trajectories and therefore reduces the minigap roughly by a factor
$l_{\text{layer}}/\delta$. This explains the strong suppression of the
minigap that sets in for values of the roughness parameter larger than
$1$ as shown in Fig.~\ref{roughinterface}.
 
A reduction of the minigap appears in the same manner, if the
superconductor is dirty, i.e., has an elastic mean free path
$l_{\text{S}}$ shorter than the coherence length
$\xi_0=v_{\text{F}}/\Delta_{\text{S}}$. The disorder close to the
interface acts similarly as the dirty layer at the interface. The role
of the ``roughness''-parameter is now taken over by
$\xi_0/l_{\text{S}}$, since Andreev reflection occurs in this layer of
the superconductor. Again the effective length of the trajectory in
the normal layer is enhanced by normal reflection at the impurities in
the superconductor. Thus, we expect a qualitative similar behavior
that the minigap will be reduced by an increase of the disorder in the
superconductor.

\section{Concluding remarks}
In conclusion, we have studied the local density of states of a
proximity sandwich for a variety of situations. At arbitrarily small
impurity concentration, a gap opens at the Fermi energy; it is maximal
if the elastic mean free path is of the order of the normal-layer
thickness. We have numerically calculated this gap and its dependence
on surface and interface roughness. Whereas the gap is relatively
stable to surface roughness, it is strongly suppressed by interface
roughness. We have also investigated the effect of a non-ideally
transmitting interface. Lastly, we have investigated the influence of
different models of impurity scattering (Born vs. unitary limit) and
shown that the two models lead to a qualitatively similar behavior.

\acknowledgements W.~B. was supported by the ``Stichting voor
Fundamenteel Onderzoek der Materie'' (FOM) and a Feodor Lynen
Fellowship of the ``Alexander von Humboldt-Stiftung''.

\appendix
\section{Numerical calculations}
\label{numerics}

For our calculations we use the Riccati parametrization of the
Eilenberger equations (see \cite{schopohl}). We represent the Green's
function on a trajectory in the form
\begin{eqnarray}
  \label{parametrization}
  \hat g=\frac{1}{1+a a^{\dagger}}\left(
  \begin{array}[c]{ll}
  1-a a^{\dagger} & 2a\\2a^\dagger & a a^{\dagger}-1 
  \end{array}\right)\,.
\end{eqnarray}
We have introduced the Andreev amplitudes $a$ and $a^\dagger$, which
depend on the variables as $\hat g$. From the Eilenberger equation
(\ref{eilenberger_equations}) one derives two decoupled numerically
stable equations of the Riccati type:
\begin{equation}
 \label{riccatiequ}
 \begin{array}{rl}
 -{\bf v_{F}\nabla}a &= 2\tilde{\omega}a + \tilde{\Delta^{\ast}}a^2 -
 \tilde{\Delta}\,,\\
{\bf v_{F}\nabla}a^{\dagger} &=
 2\tilde{\omega}a^{\dagger} + \tilde{\Delta}a^{\dagger 2} -
 \tilde{\Delta^{\ast}}\; .
\end{array}
\end{equation}
Here, the impurity self-energy is included in $\tilde{\omega}$ and
$\tilde{\Delta}$:

\begin{eqnarray}
\tilde{\omega} &=& -iE + \sigma_{11}({\bf r}) \nonumber\\
\tilde{\Delta} &=& \Delta({\bf r}) + \sigma_{12}({\bf r})\; .
\end{eqnarray}

We note that the first equation is stable in an integration in
positive direction along the trajectory, whereas the second equation
is stable for integration in the opposite direction. In the stable
direction the differential equation (\ref{riccatiequ}) is conveniently
integrated by a discretization which leads to the very accurate
expression
\begin{equation}
 \label{discretization}
 a_{n+1} = \frac{\left(\tilde{\Delta} - (\tilde{\Omega} +
 \tilde{\omega})a_n\right)
 e^{-2\tilde{\Omega} h/v_{\text{F}}} -
 \tilde{\Delta} - (\tilde{\Omega}-\tilde{\omega})a_n}
 {\left(\tilde{\Delta}a_n - \tilde{\Omega} + \tilde{\omega}\right)
 e^{-2\tilde{\Omega} h/v_{\text{F}}} -
 \tilde{\Delta}a_n - \tilde{\Omega}-\tilde{\omega}}\; .
\end{equation}
Here, $\tilde{\Omega} = (\tilde{\Delta}^2+\tilde{\omega}^2)^{1/2}$,
and $h$ is the step size.

\section{Minigap in the dirty limit}
\label{dirty minigap}

The Usadel equation \cite{usadel:70}
\begin{equation}
\label{usadel equation}
\frac{D}{2}\frac{d^2}{dx^2}\Theta = \omega \sin\Theta - 
\Delta \cos\Theta
\end{equation}
contains only one energy scale, the Thouless energy
$E_{\text{Th}}=\hbar D/d^2$ (apart from the pair potential $\Delta$).
The minigap that emerges on this scale can be explained by the
following argument. The LDOS is given by the real part of $g=\cos
\Theta$. If it vanishes, we may write $\cos \Theta = -i\sinh
\vartheta$ where $\vartheta$ is real, and the Usadel equation in the
normal metal reduces to
\begin{equation}
\label{real usadel equation}
\frac{d^2\vartheta}{d\xi^2} = \frac{2Ed^2}{D}\cosh\vartheta
\end{equation}
(here, $\xi=x/d$).  This differential equation is of elliptic type and
can be integrated twice. Next, we apply the boundary conditions
$\vartheta(\xi=0)=0$ and $\frac{d}{d\xi}\vartheta(\xi=1)=0$. The
latter accounts for the symmetry of the reflecting surface. The first
condition approximates $\vartheta$ by its value in the bulk
superconductor at low energies which is justified, if $\Delta$ is
sufficiently large.  We find that Eq.~(\ref{real usadel equation}) is
only solvable, if
\begin{equation}
E > 0.780\hbar D/d^2
\end{equation}
The minigap of the dirty limit is set by this restriction.

\end{document}